\documentclass[prd,twocolumn,showpacs,preprintnumbers,superscriptaddress,nofootinbib,floatfix]{revtex4-1}

\usepackage{braket}
\usepackage{amsfonts}
\usepackage{amsmath}
\usepackage{amssymb}
\usepackage{amsthm}
\usepackage{bm}
\usepackage{dcolumn}
\usepackage{epsfig}
\usepackage{graphicx}
\usepackage{graphics}
\usepackage{latexsym}
\usepackage{rotating}
\usepackage[colorlinks=true]{hyperref}
\usepackage[usenames]{color}
\usepackage{yfonts}
\usepackage{float}
\usepackage{ucs}
\usepackage{xspace} 
\usepackage{mathrsfs}
\usepackage{booktabs}
\usepackage[toc,page]{appendix}
\usepackage{array}
\usepackage[normalem]{ulem}
\usepackage{tikz}

\usepackage[super]{nth}

\newcommand{\sub}[1]{_{\text{#1}}}

\newcommand{\be}{\begin{equation}}
\newcommand{\ee}{\end{equation}}

\begin{document}

\title{Convergence of Fourier-domain templates for inspiraling eccentric compact binaries}

\author{Sashwat Tanay}
\email{stanay@go.olemiss.edu}
\affiliation{Department of Physics and Astronomy, The University of Mississippi, University, MS 38677, USA}

\author{Antoine Klein}
\email{antoine@star.sr.bham.ac.uk}
\affiliation{Institut d’Astrophysique de Paris, CNRS \& Sorbonne Universit\'es,
UMR 7095, 98 bis bd Arago, 75014 Paris, France}
\affiliation{School of Physics and Astronomy, University of Birmingham,
Edgbaston, Birmingham B15 2TT, United Kingdom}

\author{Emanuele Berti}
\email{berti@jhu.edu}
\affiliation{Department of Physics and Astronomy,
Johns Hopkins University, 3400 N. Charles Street, Baltimore, MD 21218, USA}

\author{Atsushi Nishizawa}
\email{anishi@kmi.nagoya-u.ac.jp}
\affiliation{Research Center for the Early Universe (RESCEU), School of Science, University of Tokyo, Bunkyo, Tokyo 113-0033, Japan}
\affiliation{Kobayashi-Maskawa Institute for the Origin of Particles and the Universe, Nagoya University, Nagoya 464-8602, Japan}

\date{\today}
	
\begin{abstract}
The space-based detector LISA may observe gravitational waves from the early inspiral of stellar-mass black hole binaries, some of which could have significant eccentricity. Current gravitational waveform templates are only valid for small orbital velocities (i.e., in a post-Newtonian expansion) and small initial eccentricity $e_0$ (``post-circular'' expansion). We conventionally define $e_0$ as the eccentricity corresponding to an orbital frequency of $5 \text{ mHz}$, and we study the convergence properties of frequency-domain inspiral templates that are accurate up to 2PN and order $e_0^6$ in eccentricity~\cite{Tanay2016}. We compute the so-called ``unfaithfulness'' between the full template and ``reduced'' templates obtained by dropping some terms in the phasing series; we investigate the conditions under which systematic errors are negligible with respect to statistical errors, and we study the convergence properties of statistical errors. In general, eccentric waveforms lead to larger statistical errors than circular waveforms due to correlations between the parameters, but the error estimates do not change significantly as long as we include terms of order $e_0^2$ or higher in the phasing.
\end{abstract}

\maketitle

\section{Introduction} 
The first two observing runs by Advanced LIGO and Virgo~\cite{LIGO-O2} detected gravitational waves (GWs) from 11 compact binary coalescence events, and the era of GW astronomy has begun.
In the near future, GW astronomy has the potential to answer several important open questions: it will test general relativity in the strong gravity regime, probe the neutron star equation of state, shed light on astrophysical formation scenarios of compact object binaries, and potentially resolve outstanding open problems in cosmology~\cite{LR_SS,2013LRR,2014LRR,Berti:2015itd,Barack:2018yly}.

This paper is motivated by the possibility of identifying the formation channels of binary black holes (BBHs) using a combination of Earth- and space-based GW detectors~\cite{Sesana:2016ljz,Wong:2018uwb,Cutler:2019krq,Gerosa:2019dbe}, and in particular by the prospect of using eccentricity measurements to distinguish between two of the main proposed formation channels~\cite{Nishizawa:2016jji,Breivik:2016ddj,Nishizawa:2016eza}: field and dynamical formation.
Most BBHs are expected to circularize by the time they enter the most sensitive band of ground based detectors, so we must rely on measurements of masses, spins, redshifts and kicks to distinguish between different formation scenarios~\cite{Gerosa:2013laa,Stevenson:2015bqa,Kovetz:2016kpi,Fishbach:2017zga,Zevin:2017evb,Barrett:2017fcw,Gerosa:2017kvu,Talbot:2018cva,Wysocki:2018mpo,Gerosa:2018wbw}. However, typical BBH eccentricities are larger for binaries formed dynamically than for binaries formed in the field (see e.g. Fig.~1 of~\cite{Nishizawa:2016eza}),
at the typical frequencies $\sim 10^{-2}$~Hz targeted by the planned Laser Interferometer Space Antenna (LISA) \cite{2013GWN.....6....4A}. Therefore LISA has the potential to measure BBH eccentricities and to shed light on their formation channel in a way that is complementary to ground based detectors \cite{Nishizawa:2016jji,Breivik:2016ddj,Nishizawa:2016eza,Samsing:2017xmd,Samsing:2018ykz,Samsing:2018isx,DOrazio:2018jnv,Samsing:2018nxk,Zevin:2018kzq,Rodriguez:2018pss}.

There is a large body of work extending the pioneering study of GWs from eccentric compact binaries by Peters and Mathews~\cite{PM63}, where the binary dynamics was treated at Newtonian order, to higher post-Newtonian (PN) orders.
The first analytic Fourier-domain templates were calculated within the stationary-phase approximation for arbitrary initial eccentricity, including the effect of periastron advance, in \cite{Mikoczi:2012qy}. Analytical expressions for the decay of the orbital parameters under radiation reaction using hypergeometric functions were derived in~\cite{1996NCimB.111..631P,Pierro:2002wd,Mikoczi:2015ewa} at Newtonian order, and in~\cite{Mikoczi:2015ewa} at 1PN order.
A Kepler-like parametrization was introduced and extended up to 3PN order in~\cite{DD85, DS88, MGS}, and the decay of the orbital parameters under radiation reaction was computed in~\cite{1985AnPhy.161...81S, 1995PhRvD..52.6882I, 2003PhRvD..68d4004K,Nissanke2004er}. Now the evolution of the orbital parameters under radiation reaction is known up to 3.5 PN order~\cite{DGI,konigs}. A time-domain template for compact binaries where the orbital elements were evolved via numerical integration was introduced in Ref.~\cite{Tanay2016}, and it can be regarded as the eccentric extension of the popular \texttt{TaylorT4} approximant for quasicircular binaries. 

In the analysis of quasicircular binary inspirals it is common to use analytic Fourier-domain templates computed within the stationary-phase approximation (SPA), such as the \texttt{TaylorF2} template. Reference~\cite{YABW09} generalized these Fourier-domain templates to eccentric binaries, computing templates which are valid at Newtonian order and up to order $e_0^8$ in a small-eccentricity expansion of the phase (here $e_0$ is defined to be the eccentricity at some reference orbital frequency $f_{\rm orb}=f_{\rm orb,\,0}$).

This work was extended to 2PN-$e_0^6$ in Ref.~\cite{Tanay2016}, which will be the starting point of our study. Moore et al.~\cite{2016PhRvD..93l4061M} extended Ref.~\cite{Tanay2016} to 3PN order, but only at leading order in $e_0$. All of the templates above are valid at Newtonian order in amplitude and do not include the effect of periastron advance in the phase. Reference~\cite{2018CQGravityYunes} constructed analytic templates which are valid for arbitrary initial eccentricity $e_0$ within the SPA using a truncated sum of harmonics and hypergeometric functions. This work was recently extended to 3PN accuracy~\cite{2019arXiv190305203M}, and efforts are underway to extend Ref.~\cite{YABW09} up to 3PN-$e_0^6$ order in phase and 1PN order in amplitude, incorporating periastron advance effects~\cite{STiwari2019}.

With so many parallel efforts on analytic Fourier-domain eccentric templates underway, it is crucial to investigate the convergence of these proposed GW templates.
This is the main goal of our work. We study the convergence properties of the 2PN-$e_0^6$ accurate template proposed in Ref.~\cite{Tanay2016}.
This ``\textit{fiducial template}'' is a sum over harmonics (labeled by $j$), where each harmonic has a phase which itself is a bivariate series in the initial eccentricity $e_0$ and in the PN parameter
\be
x \equiv \left(\frac{2\pi G m_z f_{\rm orb}}{c^3}\right)^{2/3}\,,
\ee
where $f_{\rm orb}$ is the orbital frequency for a circular binary, and $m_z=(1+z)m$  is the redshifted total mass of the binary. A preliminary, more limited investigation of the convergence of this bivariate series in the context of parameter estimation can be found in~\cite{Nishizawa2016}. Our work should be helpful in guiding future efforts to extend the above templates to higher orders, and it can readily be generalized as soon as more accurate templates become available.

We focus on the convergence of the expansion of the phasing (rather than the amplitude) because the phasing is known to have greater impact on detectability and parameter estimation. We first drop some terms from the fiducial template to get (presumably) less accurate ``reduced'' templates, then we perform calculations of the so-called ``unfaithfulness'' between these reduced templates and the fiducial 2PN-$e_0^6$ accurate template to assess the importance of the dropped term(s). We also investigate the conditions under which systematic errors due to dropping high-order terms from the fiducial template exceed the statistical errors. This is useful because whenever systematic errors are negligible with respect to statistical errors, one can choose a truncated template to improve computational efficiency in parameter estimation. Finally we study the convergence of  statistical errors.

The paper is organized as follows. Section~\ref{sectionI} describes the waveform model and gives details on our calculation of matches and Fisher matrices. Section~\ref{dabg} is an overview of data analysis concepts that are relevant for our study. Our results on unfaithfulness are presented in Sec.~\ref{subsection_mismatch}, the comparison of systematic and statistical errors is shown in Sec.~\ref{systematic vs statistical}, and the convergence of statistical errors is studied in Sec.~\ref{statistical error subsection}. In Sec.~\ref{conclusions} we summarize our main results and outline directions for future work.  To improve readability, some technical details are relegated to the Appendices. Appendix~\ref{appendix_A} illustrates why certain cross terms in the Fisher matrix integrands can be neglected due to their oscillatory nature, and Appendix~\ref{beam_pattern_appendix} defines beam pattern functions and other quantities appearing in the calculation of the GW strain. The code used in our analysis is publicly available online~\cite{EccentricPNwaveform}.

\section{Waveforms, match and Fisher matrix calculations}
\label{sectionI}

In this section we describe our waveform model, and then we give details of our unfaithfulness and Fisher matrix calculations. 
 
Our analytic frequency-domain GW template for compact binaries inspiraling in eccentric orbits uses Newtonian amplitudes, the first six harmonics $(j =1,\dots,6)$, and a 2PN-$e_0^6$ accurate phase. Here $e_0$ is the eccentricity at which the orbital frequency of the binary is $f_0/2$, and we (somewhat arbitrarily) set $f_0= 10$ mHz. In other words, when the binary has eccentricity $e_0$, the second harmonic (which dominates the signal for small eccentricities) has frequency $f_0$. From Appendix~\ref{beam_pattern_appendix} and Eq.~(3.11) of \cite{YABW09} it follows that if we include six harmonics, the expression of $e$ used in the amplitude should be accurate up to ${\cal}{O}(e_0^3)$. Below we list the relevant expressions only at Newtonian order for illustration, but in the actual calculations we retained all terms up to 2PN-$e_0^6$ order; these can be found in Appendix A of~\cite{Tanay2016}. The waveform in the stationary phase approximation has the form
\begin{widetext}
\begin{equation} 
\tilde{h}(f) =    \frac{ \sqrt{3}}{2}  \mathcal{\tilde{A}}    {\left(\frac{G m_z \pi f}{c^3}\right)}^{-7/6}     \sum\limits_{j=1}^{6} \xi(f,j)
{\left(\frac{j}{2}\right)}^{2/3}  e^{-i(\pi/4 + \Psi (f,j) -  \phi_D (f,j))} \,,             \label{GW strain}       
\end{equation}
where the symmetric mass ratio $\eta=m_1 m_2/m^2$,
\begin{subequations}                       \label{GW strain2}                   
\begin{align}
 \mathcal{\tilde{A}} &=  - {\left(\frac{5 \eta \pi}{384}\right)}^{1/2}  \frac{G^2 m_z^2}{c^5 D_L}  ,  \\             
 \xi(f,j)  &=     \frac{\left(  1-e(f,j)^2\right)^{7/4}}{{\left( 1+\frac{73}{24}e(f,j)^2+\frac{37}{96}e(f,j)^4\right)}^{1/2}}    \left(   \Gamma(f,j)  +i~ \Sigma(f,j)   \right)     \,  ,                                                     
\end{align}
\end{subequations}
the quantity $D_L$ is the luminosity distance to the source, and 
\begin{align}                          
 \Gamma(f,j) =     F_{+}(f,j) C_{+}^{j}(f)  +  F_{\times}(f,j) C_{\times}^{j}(f)\,,                 \label{GW strain3} \\           
 \Sigma(f,j)  =    F_{+}(f,j) S_{+}^{j}(f)  +  F_{\times}(f,j) S_{\times}^{j}(f)\,.                  \label{GW strain4}
\end{align}
The definitions of $F_{+}, F_{\times}, C_{+}^{j}, C_{\times}^{j}, S_{+}^{j}, S_{\times}^{j}$ are given in Appendix~\ref{beam_pattern_appendix}.    
The Fourier phase $\Psi(f,j)$ and $e(f,j)$, up to the leading PN order and sixth order in $e_0$, are given by
\begin{align}               
  \Psi (f,j) &=  
    j \phi_c - 2\pi f t_c - \frac{3}{128 \eta} \left(  \frac{G m_z \pi f}{c^{3}}\right)^{-5/3}        \left(\frac{j}{2}\right)^{8/3}      \mathcal{C}(f,j),       \label{Psi_eqn}   \\
     e (f,j) &=   e_0\chi(f,j)^{-19/18} + \frac{3323}{1824} e_0^3 \left(\chi(f,j)^{-19/18}- \chi(f,j)^{-19/6}\right) + 
\left( \frac{15 994 231
}{6 653 952}\chi(f,j)^{-19/18}   \right.\nonumber\\      
&\qquad \left.       -\frac{11 042 329
}{1 108 992} \chi(f,j)^{-19/6  }                      
        +\frac{50 259 743
}{6 653 952} \chi(f,j)^{-95/18}\right)e_0^5 ,             \label{e_eqn}                             
\end{align}
with $\chi(f,j)= 2 f/(j f_0)$ and
\begin{align}
 &\mathcal{C}  (f,j) = 1- \frac{2355}{1462}e_0^{2}~\chi(f,j)^{-19/9} + \left( -  \frac{2608555}{444448}\chi(f,j)^{-19/9}+ \frac{5222765}{998944}\chi(f,j)^{-38/9} \right)e_0^{4}  \nonumber\\ 
  &\qquad {}+ \left(- \frac{1326481225}{10134144}\chi(f,j)^{-19/9}  + \frac{173355248095}{455518464}\chi(f,j)^{-38/9}- \frac{75356125}{3326976}\chi(f,j)^{-19/3}  \right)e_0^{6}.   \label{psi_C_eqn} 
\end{align}
We also introduced the Doppler phase
\begin{align}
\phi_D(f,j)  =  2 \pi R f \sin \bar{\theta}_S  \cos\left( \bar{\phi}(f,j) - \bar{\phi}_S  \right),                   \label{phi_D_eqn} 
\end{align}
where $R = 1 AU$ and the orbital phase $\bar{\phi}(f,j)$ of LISA's barycenter around the Sun is
\begin{align}
&  \bar{\phi}(f,j) =\frac{ 2 \pi t(f,j)}{\mathrm{T_0}} = \frac{1}{ \mathrm{T_0}} \left[ 2 \pi t_c  + \frac{G m_z}{\eta c^3}    \left( \frac{2 G m_z \pi f}{j c^3} \right)^{-8/3}   \left(  - \frac{5}{256} + \frac{785 e_0^2}{11008 } \chi(f,j)^{-19/9} +   e_0^4  \left(  -\frac{5222765}{14475264} \chi(f,j)^{-38/9}   \right.\right.\right.\nonumber\\ 
  &\qquad \left.\left.\left.{}    +\frac{2608555}{10039296} \chi(f,j)^{-19/9}   \right)       +   e_0^6  \left(  \frac{75356125}{35487744} \chi(f,j)^{-19/3}  +\frac{17355248095}{6600720384} \chi(f,j)^{-38/9}  +\frac{1326481225}{2288959488} \chi(f,j)^{19/9}   \right)         \right)    \right].                        \label{phi_bar_eqn} 
\end{align}
\end{widetext} 
Here $T_0=1$~yr, the angles $  (  \bar{\theta}_S , \bar{\phi}_S    )  $ define the direction of the source in the Solar barycenter frame, and $t_c$ and $\phi_c$ denote the time and phase at coalescence, respectively~\cite{Berti:2004bd,Yagi2010}. The amplitudes $C_{+}^{j}, C_{\times}^{j}, S_{+}^{j}, S_{\times}^{j}$ are computed by keeping the first two terms in the expansion of Eq.~(\ref{e_eqn}): cf. Appendix~\ref{beam_pattern_appendix}. This is consistent with retaining a Newtonian amplitude and six harmonics. The beam pattern functions $F_{+}$ and $F_{\times}$ depend on $f$ and $j$ through $\bar{\phi}(f,j)$. The equations listed above are of order $x^0$  in a PN expansion, hence they do not depend on $x$.

\begin{table*}[htp]
\begin{center}
\centering
\begin{tabular}{| *5{>{\centering\arraybackslash}m{1in}|} @{}m{0pt}@{}}
\hline
 \centering   {2PN ($x^2$) }     &     A   &       B   &    C   &         D     \\ 
\hline
 \centering       {1.5PN  ($x^{3/2}$)  }     &      &      &       &   E   \\ 
\hline
\centering    {1PN  ($x^1$) }   &      &      &       &   F    \\ 
\hline
 \centering   {Newtonian  ($x^0$)}  &      &      &       &   G   \\ 
\hline
        & \centering    {$e_0^0$      }       &       {      $e_0^2$    }         &      {       $e_0^4$    }       &        {      $e_0^6$    }      \\   
\hline
\end{tabular}
\end{center}
\caption{Template naming conventions. According to the alphabetical naming convention, the template obtained by dropping terms corresponding to the letters B, C and D from the phase is called ``template B'', and so on (see text). In the curly bracket convention, the $ \left\lbrace  e_0^2 \right\rbrace$ template is obtained by retaining terms of order up to $e_0^2$ (i.e., the two leftmost columns). Additionally, ``template H'' corresponds to dropping cells C, D and E from the fiducial template.}
\label{define templates}
\end{table*}

The above template is slightly modified with respect to Ref.~\cite{Tanay2016}: the amplitude has an extra factor of $\sqrt{3}/2$ to account for the $60^{\circ}$ opening angle of the LISA arms~\cite{berti2005estimating}, and the Fourier phase has been changed from $\Psi$ to $(\Psi -\phi_D)$ to account for the Doppler phase due to the motion of the detector around the Sun~\cite{Nishizawa2016,Yagi2010}.

\subsection{Fiducial template and truncated templates}
\label{subsec:truncate}

We will now introduce the structure of the templates used in our calculations. We refer to the 2PN-$e_0^6$ order accurate template of~\cite{Tanay2016} as the ``\textit{fiducial template}''. To assess convergence, we also consider various ``truncated templates,'' i.e. templates derived from the fiducial one by dropping certain terms in the phase. The amplitude of all templates is accurate at Newtonian order and ${\cal{O} }(e^4)$, because the phase plays a more important role than the amplitude for detection and parameter estimation (see e.g. \cite{DGI,konigs,YABW09,Tanay2016}). 

Table~\ref{define templates} illustrates the difference between the various templates. Each cell in the table represents a term of a certain order in the PN frequency parameter $x$ (rows) and in the initial eccentricity $e_0$ (columns). We will use two different notations to distinguish between templates.

A ``letter-based'' template means that we drop all terms of order greater than or equal to the corresponding cell in the table. For example, ``waveform A'' is obtained by neglecting cells A, B, C and D, i.e. all of the 2PN corrections to the waveform; ``waveform B'' is obtained by neglecting cells B, C and D, i.e. all 2PN corrections of order $e_0^2$ and higher in the initial eccentricity; and ``waveform C'' is obtained by neglecting cells C and D. Similarly, ``waveform G'' is obtained by neglecting cells G, F, E and D; ``waveform F'' is obtained by neglecting cells F, E and D; and ``waveform E'' is obtained by neglecting cells E and D. ``Waveform D'' corresponds to neglecting only the 2PN, $e_0^6$ term.

We will also use a curly bracket notation $ \left\lbrace  y^n \right\rbrace$, meaning that the phase is $y^n$ accurate in the parameter $y$, where $y$ stands either for $e_0$ or for $x$ in the bivariate series for the Fourier phase of Eq.~(\ref{Psi_eqn}). For example, the $\left\lbrace  e_0^2 \right\rbrace$ template is accurate up to order $e_0^2$ (and 2PN) in phase, i.e. we retain the first two columns from the left in Table~\ref{define templates}. Likewise, for the $ \left\lbrace  x^1 \right\rbrace$ template we retain terms up to 1PN (and order $e_0^6$) in phase, i.e. the two bottom rows in Table~\ref{define templates}.

The fiducial template of Eq.~(\ref{GW strain}) is a series of harmonics labeled by the integer $j$, where the phase $\Psi(f,j)$ of each harmonic is itself a bivariate series in $x$ and $e_0$: cf. Eqs.~(\ref{Psi_eqn}) and (\ref{psi_C_eqn}). Here we focus on the convergence of $\Psi(f,j)$ as a bivariate series because, as already mentioned, the phase of a GW template is more important than the amplitude (as long as $e_0$ is small, so that a small-$e_0$ expansion is valid). \footnote{At Newtonian order, the radiation reaction timescale $T\sub{rr} = \omega/\dot{\omega}$ (where $\omega=2\pi f_{\rm orb}$ is the angular frequency) is
\begin{align}
T\sub{rr} = \frac{5 G{\cal{M}}}{96 c^3}  \left( {\frac{G \cal{M} \omega}{c^3}}\right)^{-8/3} \left[  \frac{ \left( 1-e^2\right)^{7/2} }{ 1 + \frac{73}{24}e^2  +\frac{37}{96}e^4 }\right].\nonumber
\end{align}
The quantity in square brackets -- say, $Z(e)$ -- must be expanded for small $e$ to finally arrive at the expression of the Fourier phase $\Psi$ which occurs in Eq.~(\ref{GW strain})~\cite{YABW09}. Any Taylor series has a radius of convergence equal at most to the distance from the expansion point (here $e=0$) and the nearest singularity in the complex plane~\cite{arfkenWeber}, which here is located at $e\sim \pm ~0.58~i $. Therefore none of our templates should be trusted beyond $e_0 \sim 0.58$ (although they may become unfaithful for much smaller values of $e_0$).} In this work we have retained only the first six harmonics $(j=1,\dots,6)$ in all of our templates. The convergence of the harmonic expansion is an interesting topic for future work. 

\subsection{Match calculations}

Our convergence analysis of PN, small-eccentricity waveforms is based on some data analysis concepts that we introduce below. First of all, we define the ``faithfulness'' $M$ between two GW signals $h_1(t)$ and $h_2(t)$ as the following integral, maximized over the time and phase of coalescence  $t_c$ and $\phi_c$:
\begin{align}
M  & =      \max_{t_c, \phi_c}    \frac{ (h_1,h_2)    }{\sqrt{     (h_1,h_1)  (h_2, h_2)           }}     ,
\end{align} 
and the ``unfaithfulness'' as $(1-M)$. 
The inner product between two waveforms  $(h_1, h_2)$ is defined as
\begin{align}
\label{Eq:innerproduct}
 (h_1,h_2)    &= 4\,   \Re   \,  \int_{f_{\rm min}}^{f_{\rm max}} \, 
\frac{\tilde h_1^*(f)\, \tilde h_2(f)}{S_{\rm h}(f)} df \,,
\end{align}
where $\tilde h_i(f)$ stands for the Fourier transform of $h_i(t)$ and $S_h(f)$ is the noise power spectral density of the LISA detector~\cite{robsoncornish}.
The signal-to-noise ratio (SNR) $\rho$ of template $h$ in the detector can be estimated by
\begin{align}
 \rho^2(h) &= (h | h).
\end{align}

Unlike~\cite{robsoncornish}, we do not use a sky-averaged response, and thus we should not include the corresponding factor of 5 in the noise curve. We also treat the two channels separately (so our noise curve differs by an extra factor of 2) and we include the geometrical factor $\sqrt{3}/2$ in the waveform definition~\eqref{GW strain}, yielding an extra factor of 3/4. For these reasons, we use the noise curve of~\cite{robsoncornish} {\em without} the overall factor 10/3, i.e.
\begin{align}
S_h(f)& =  \frac{1}{L^2} \left[  P_{\mathrm{OMS}}(f)  +   \frac{4 P_{\mathrm{acc}}(f)}{(2 \pi f)^4}   \right] \left[ 1 + \frac{6}{10} \left(\frac{f}{f_{*}} \right)^2   \right]   \nonumber \\
&+ S_c(f),
\end{align}
where $L = 2.5 ~\mathrm{Gm}$ and $f_{*} = 19.09 ~\mathrm{mHz}$. Furthermore $P_{\mathrm{OMS}}(f)$, $P_{\mathrm{acc}}$ and the confusion noise $S_c(f)$ are given by
\begin{align}
P_{\mathrm{OMS}}   &= (1.5 \times 10^{-11} \mathrm{m})^2 \left[ 1+ \left( \frac{2 \mathrm{mHz}}{f} \right)^4 \right] \mathrm{Hz}^{-1},     \\
P_{\mathrm{acc}}  &= (3\times 10^{-15} \mathrm{m ~s^{-2}})^2 \left[ 1 +  \left(\frac{0.4~\mathrm{mHz}}{f}\right)^2 \right]     \nonumber  \\
&  \left[  1 +  \left( \frac{f}{8~\mathrm{mHz}} \right)^4  \right] \mathrm{Hz^{-1}}, \\
S_c(f) &= A  \left(\frac{f}{\mathrm{Hz}}\right)^{-7/3} e^{-f \alpha + \beta f \sin (\kappa f)} \times \nonumber \\ &\left[ 1 + \tanh(\gamma (f_k-f)) \right] \mathrm{Hz}^{-1},
\end{align}
where $A = 9\times 10^{-45}$ and all parameters have been chosen corresponding to an observation time of 2 years: $\alpha = 0.165~ \mathrm{Hz}^{-1},~\beta = 299~ \mathrm{Hz}^{-1}, ~\kappa = 611~ \mathrm{Hz}^{-1}, ~\gamma = 1340~ \mathrm{Hz}^{-1}$ and $f_k = 0.00173 ~\mathrm{Hz}$.

\subsection{Fisher matrix and cross terms}
Our eccentric GW templates depend on 11 parameters: $\ln {\cal{M}}_z, \ln \eta, t_c, \phi_c, \ln {D_L}, e_0, \bar{\theta}_S, \bar{\phi}_S,\bar{\theta}_L, \bar{\phi}_L$ and $\beta$. Here ${\cal{M}}_z=\eta^{3/5} m_z$ is the redshifted chirp mass, $\eta=m_1 m_2/m^2$ is the symmetric mass ratio, $t_c$ and $\phi_c$ are the coalescence time and orbital phase, $D_L$ is the luminosity distance and $e_0$ is the initial eccentricity, defined as the eccentricity corresponding to an orbital frequency $f_{\rm orb}=5$~mHz (so that the second harmonic of the radiation is at $10$~mHz). The angles $  (  \bar{\theta}_S , \bar{\phi}_S    ) $ and the angles $(\bar{\theta}_L,\bar{\phi}_L)$ define the direction of the source and of the binary's orbital angular momentum in the Solar barycenter frame. Finally, $\beta$ represents the position of the pericenter in the binary's orbital frame~\cite{YABW09, MP99}.

Let ${\bf p}=\{p_A\}$ be a vector whose components are any of these eleven parameters. The Fisher matrix is defined as the matrix with elements
\begin{align}
\tau_{AB} = 4  \Re{  \int_{0}^{\infty}   \left[ {\frac{\partial \tilde{h}(f)}{\partial p_A}}^{\ast}    \frac{\partial \tilde{h}(f)}{\partial p_B}   \right]    \frac{1}{S_h(f)} df   }. \label{cfm1}
\end{align}
In the high SNR regime, the statistical errors associated with estimating the parameters can be approximated by the square root of the diagonal elements of the inverse Fisher matrix~\cite{Vallisneri:2007ev}.

Each of our templates $\tilde{h}(f)$ is obtained by summing over the first six harmonics, i.e. $\tilde{h}(f) = \sum_{j=1}^6 \tilde{h}_j$. Therefore the integrand above contains 36 terms: 6 ``diagonal terms'' where both harmonics are the same, and 30 ``cross terms'' involving different harmonics. Examples of  a diagonal term and of a cross term are 
\begin{align}
4  \Re{  \int_{0}^{\infty}   \left[ {\frac{\partial \tilde{h}_1(f)}{\partial p_A}}^{\ast}   \frac{\partial \tilde{h}_1(f)}{\partial p_B}   \right]    \frac{1}{S_h(f)} df   },
\end{align}
and
\begin{align}
4  \Re{  \int_{0}^{\infty}   \left[ {\frac{\partial \tilde{h}_2(f)}{\partial p_A}}^{\ast}   \frac{\partial \tilde{h}_3(f)}{\partial p_B}   \right]    \frac{1}{S_h(f)} df   }   ,          \label{cfm2}
\end{align}
respectively. 

Numerical calculations show that the cross terms oscillate rapidly and that they do not significantly contribute to the integral, so they can be dropped.
An analytical justification for this approximation can be found in Appendix~\ref{appendix_A}.

\subsection{Binary catalog and cosmology}

To investigate the statistical properties of our waveforms we use a catalog of $1000$ systems. The individual source-frame masses of the binary components are uniformly distributed between $5 $\(\textup{M}_\odot\) and $45 $\(\textup{M}_\odot\). The angles $\bar{\theta}_S, \bar{\phi}_S,\bar{\theta}_L, \bar{\phi}_L $ are uniformly distributed over the sphere, and the angle $\beta$ is uniformly distributed between $[0, 2 \pi]$. We set $t_c$ and $\phi_c$ equal to $0$. We truncate the Fisher matrix integrals of Eq.~(\ref{cfm1}) at $f_{\rm max}=1$~Hz, and we choose $f_{\rm min}$ so that the observation time is 2 years. We assume a $\Lambda$CDM cosmology and a spatially flat universe
with $H_0 = 67.36$ km/s/Mpc, $\Omega_{M} = 0.3153$ and $\Omega_{\Lambda} = 0.6847$ and we fix $z=0.1$ (corresponding to $D_L = 447.8$~Mpc) for all binaries.

\section{Some data analysis background and motivation}
\label{dabg}

There are two kinds of parameter estimation errors: systematic and statistical. Systematic errors are due (e.g.) to mismodeling of GW signals, and statistical errors are due to the noise in ``ideal'' detectors (real detectors usually contribute to systematic errors as well)~\cite{creighton2012gravitational}. If the systematic errors associated with neglecting some terms in the waveform template are smaller than statistical errors, we can safely neglect those terms and increase the computational efficiency of parameter recovery without compromising its accuracy. Here we follow Appendix G of~\cite{chatziioannou2017constructing} and we introduce a criterion to decide whether systematic errors are smaller than statistical errors.

Consider a detection scenario where systematic errors are negligible, so all parameter estimation errors are statistical and due to noise. Assume also that the SNR for this detection is large enough that the posterior probability distribution is sharply peaked close to the true parameter values $\bm{p}_0$ \cite{creighton2012gravitational}.
For a parameter vector $\bm{p}$ close to $\bm{p}_0$, the unfaithfulness
is $1-M_{\mathrm{sta}}$, and the mean value of the unfaithfulness over the posterior probability is~\cite{chatziioannou2017constructing}
\begin{align}
1-E(M_{\mathrm{sta}}) = \frac{ (D-1)}{2 ~\mathrm{SNR}^2},   \label{mismatch-SNR}
\end{align}
where $E$ denotes the expectation value, $D=11$ is the dimension of our parameter space, and the subscript ``sta'' stands for ``statistical.''

Let us now include systematic errors. If we demand systematic errors to be negligible with respect to statistical errors, then the unfaithfulness $(1-M)$ due to GW mismodeling should be negligible with respect to the expected value of the unfaithfulness in Eq.~(\ref{mismatch-SNR}), i.e.
\begin{align}
1-M    \ll \frac{ (D-1)}{2 ~\mathrm{SNR}^2}.   \label{mismatch-SNR-2}
\end{align}

We also need a measure of the overall statistical uncertainty. Following e.g. Lyons~\cite{lyons}, we can associate an $n$-dimensional error ellipsoid (about the maximum of the posterior distribution) with a Gaussian posterior in an $n$-dimensional parameter space. This ellipsoid is the region of $1~\sigma$ confidence interval, within which the parameter vector can be found with probability $\sim 0.68$. The square roots of the diagonal elements of the inverse Fisher matrix are the projections of this error ellipse on the parameter axes. Define $\epsilon$ as the product of Fisher errors on all parameters, and $\epsilon_0$ as the volume of the error ellipsoid. It can be shown that $\epsilon_0$ is given by the square root of the determinant of the inverse Fisher matrix and that $\epsilon \geq  \epsilon_0$, where the equality is realized when the parameters are uncorrelated with each other, while having $\epsilon /\epsilon_0>1$ means that there is some correlation between the parameters. The parameter $\epsilon_0$ can be considered an overall measure of statistical errors and it is unaffected by a linear transformation of the parameters (or equivalently, by a rotation of the parameter axes): in fact, $\epsilon_0$ measures the volume of the error ellipsoid, which should be independent of the orientation of the parameter axes. The ratio $ \epsilon/\epsilon_0$ (``correlation factor'') quantifies the degree of correlation among the parameters, or the misalignment of the error ellipsoid with the parameter axes.

\begin{figure}[h]
\begin{center}
\includegraphics[width=0.4\textwidth, angle=0]{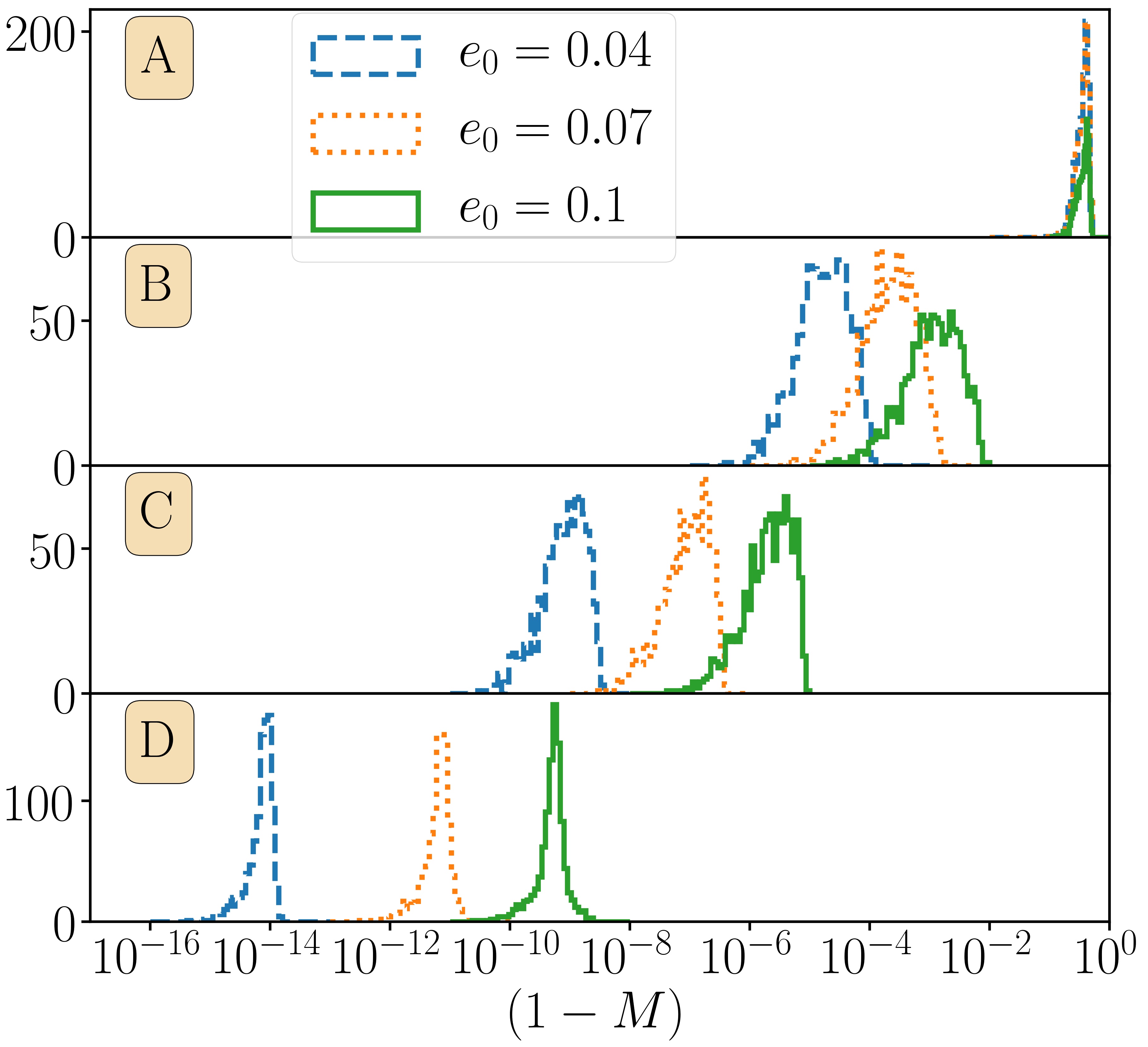}\\
\includegraphics[width=0.4\textwidth, angle=0]{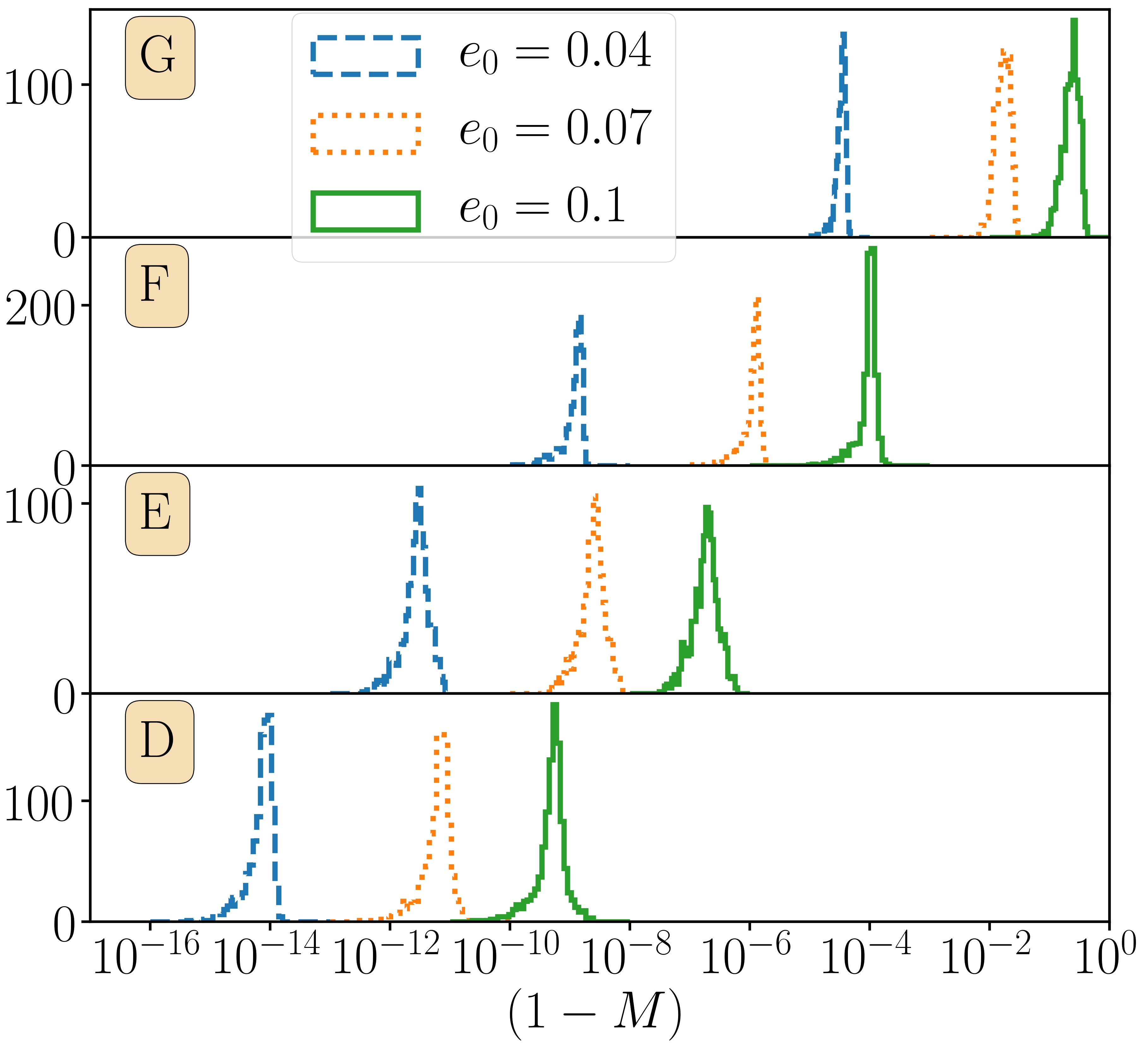}\\
\includegraphics[width=0.4\textwidth, angle=0]{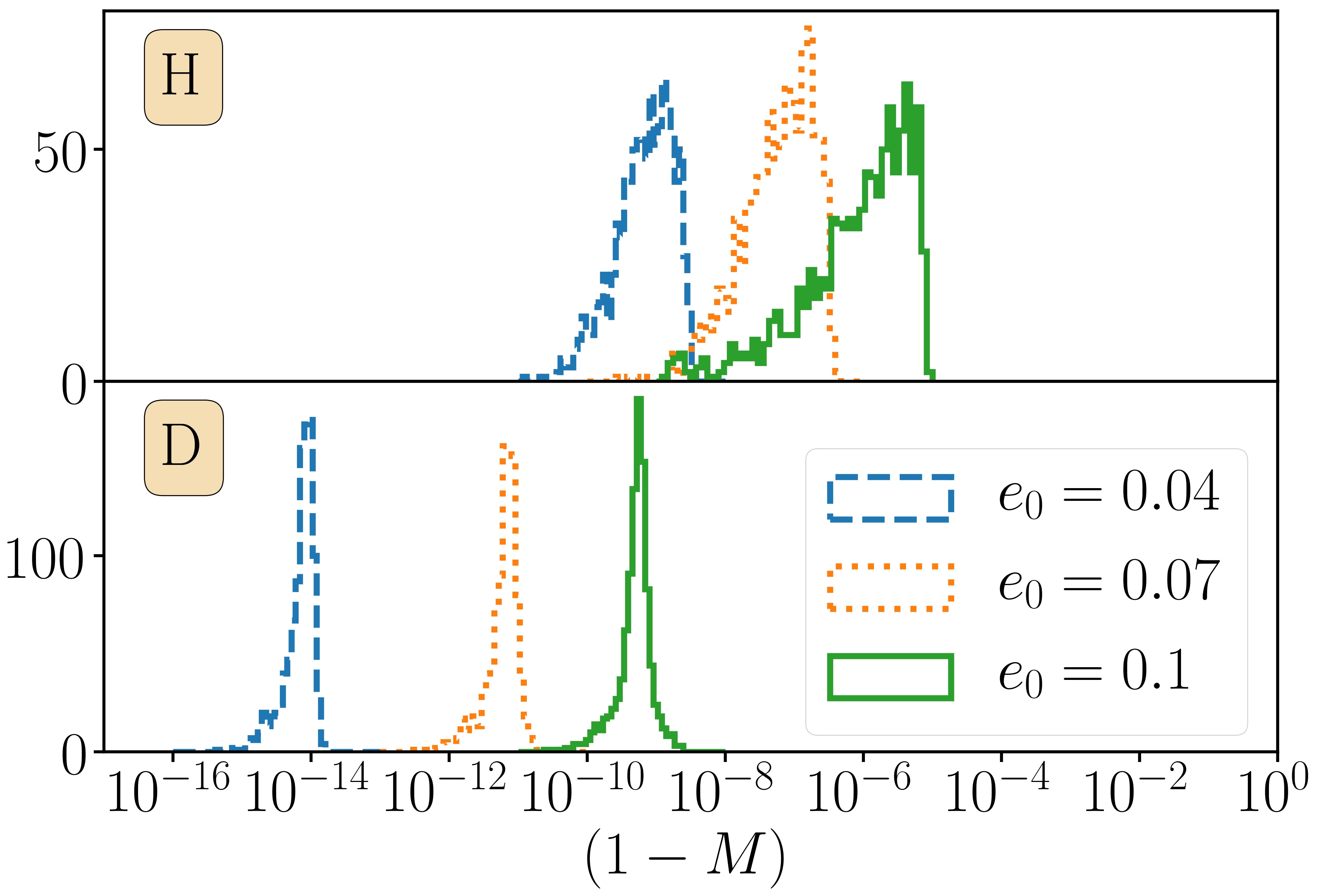}
\end{center}
\caption{
\label{mismatch_histograms}
Unfaithfulness histograms for templates A-D (top), D-G (middle), D and H (bottom) with respect to the fiducial template for three selected values of $e_0$.
See the discussion around Table~\ref{define templates} for a definition of the templates.
} 
\end{figure}

\begin{figure*}[t]
\begin{center}
\includegraphics[width=0.45\textwidth, angle=0]{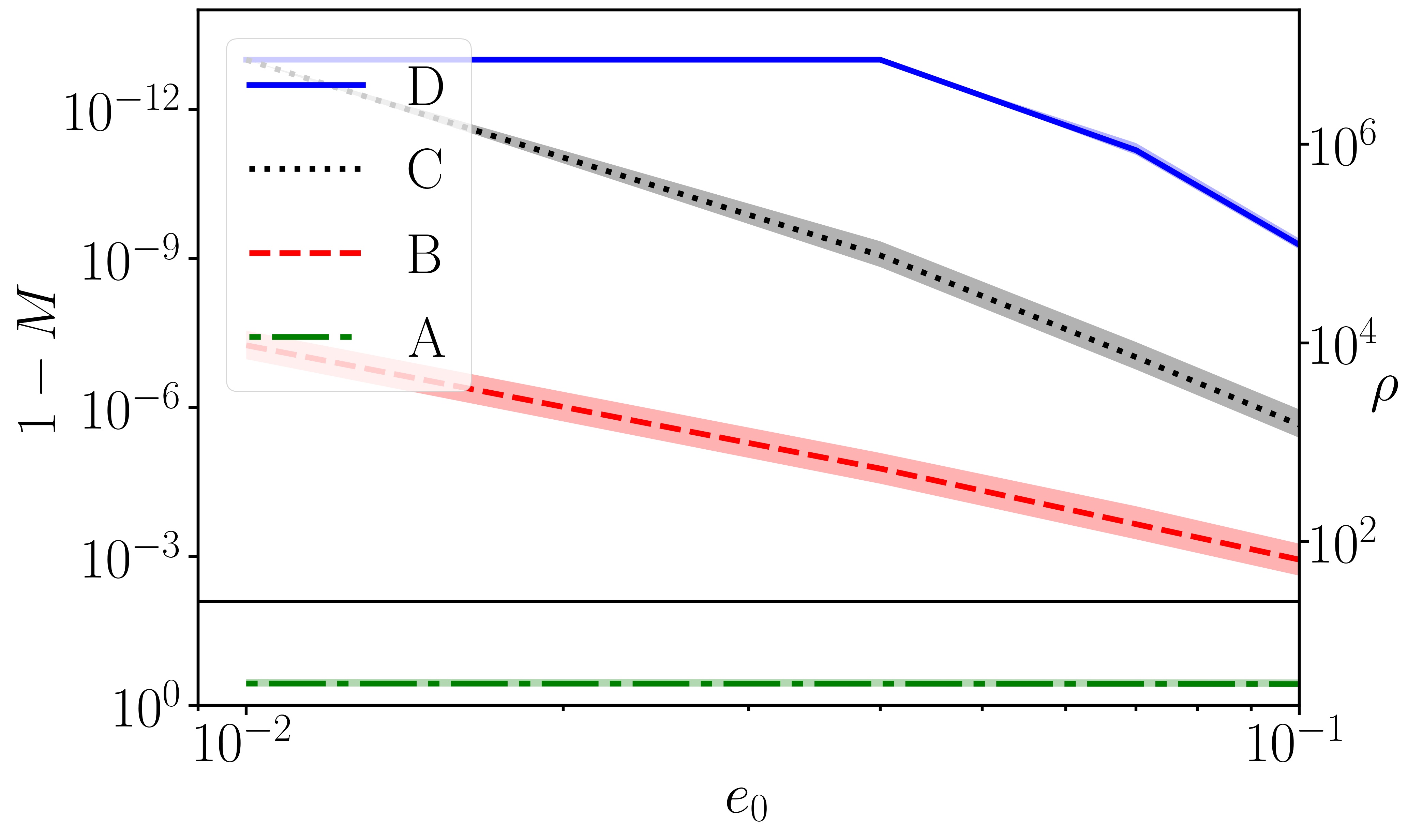}
\includegraphics[width=0.45\textwidth, angle=0]{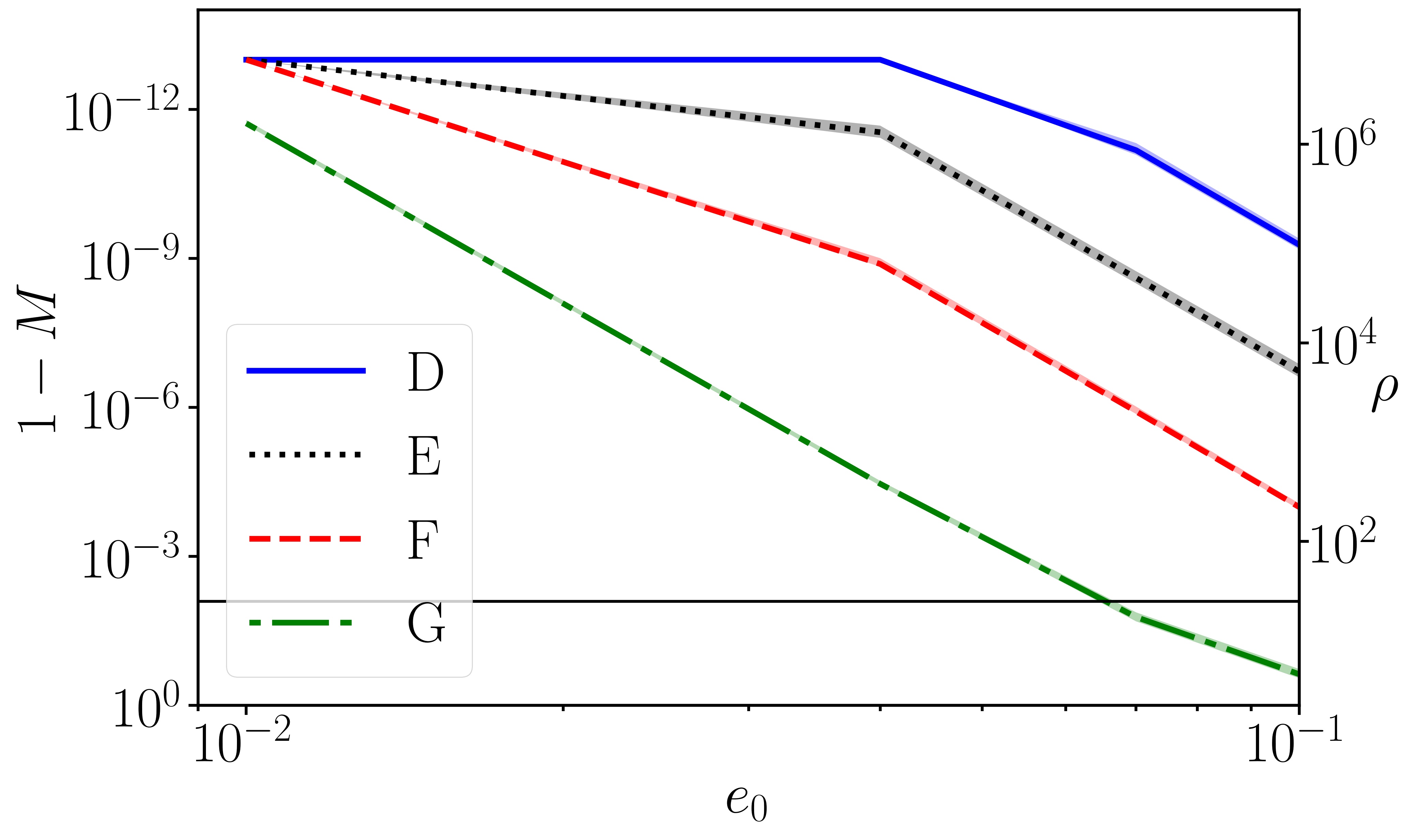}\\
\includegraphics[width=0.45\textwidth, angle=0]{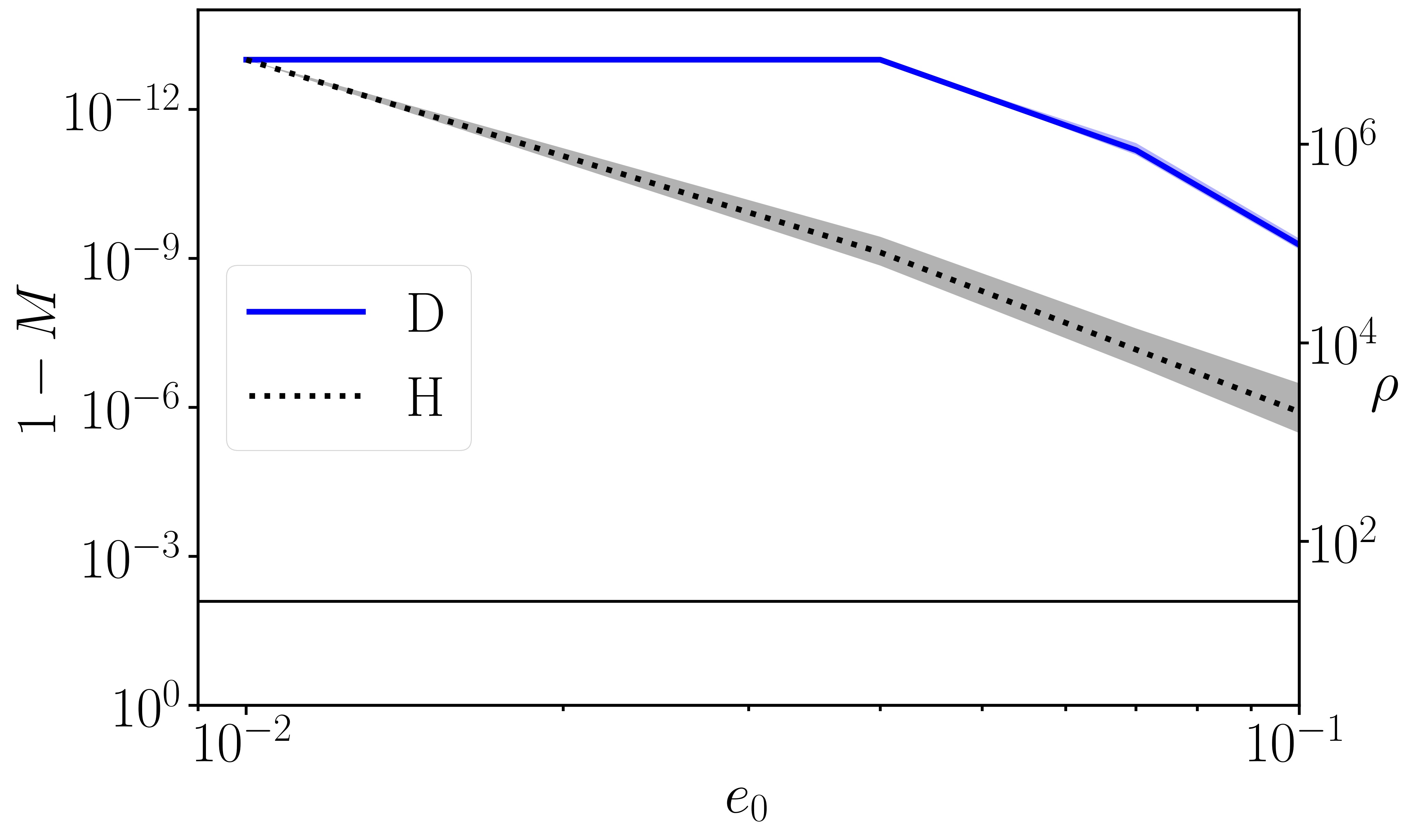}
\includegraphics[width=0.45\textwidth, angle=0]{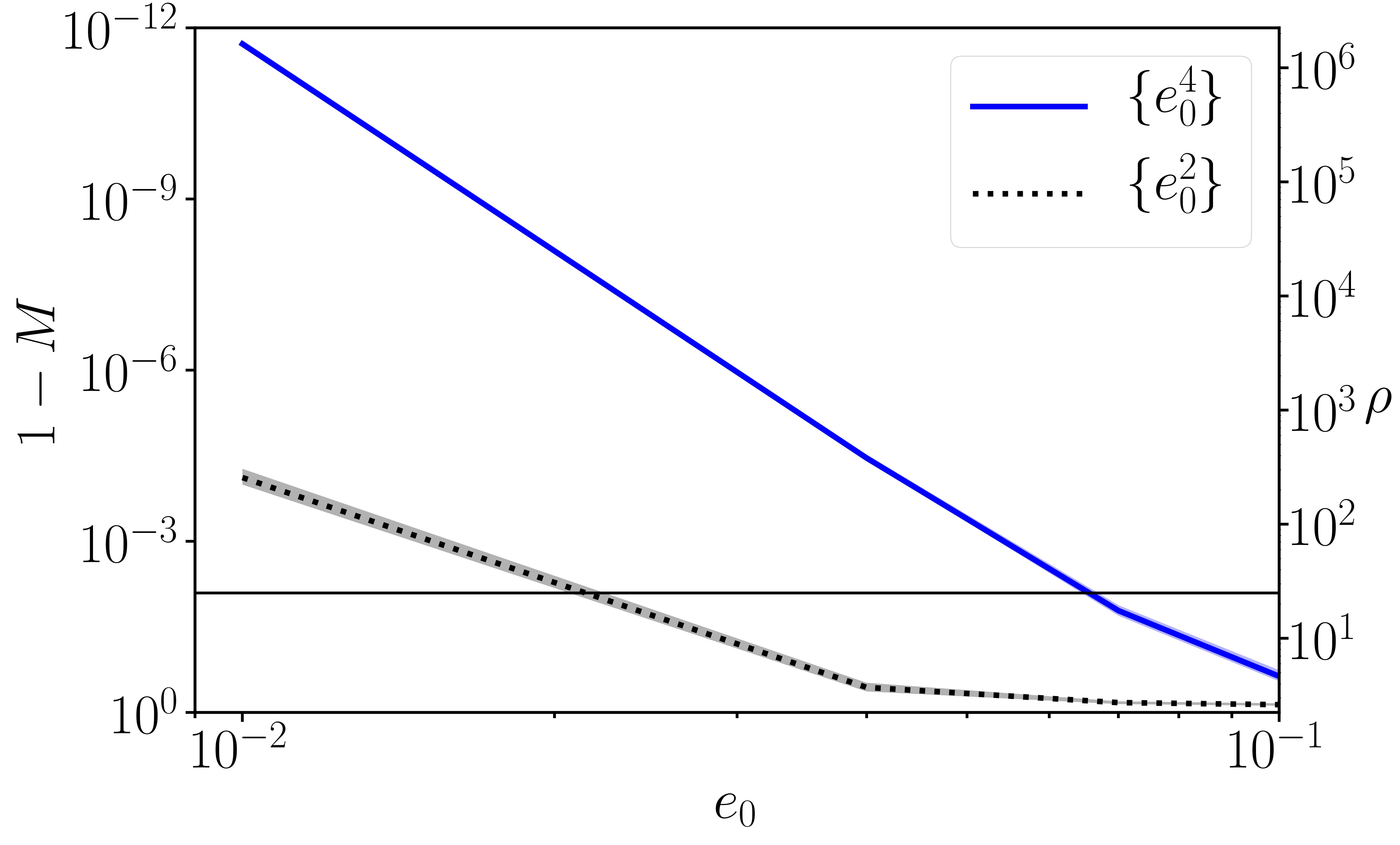}
\end{center}
\caption{
\label{mismatch_SNR}
Unfaithfulness-SNR plots for various templates with respect to the fiducial template. Top left: templates A-D; top right: templates D-G; bottom left: templates D and H; bottom right: templates $ \left\lbrace e_0^4  \right\rbrace$ and $ \left\lbrace e_0^2  \right\rbrace$. The SNR on the secondary y-axis corresponds to the equal sign in Eq.~(\ref{mismatch-SNR-2}).
Lines correspond to the median unfaithfulness (SNR) for our sample of 1000 binaries, and shaded regions correspond to the 25th and 75th percentiles.
} 
\end{figure*}

\section{Results}
\label{resultsSection}

In this section we present our results on the convergence properties of the 2PN-${e_0}^6$ accurate bivariate template (the fiducial template) proposed in Ref. \cite{Tanay2016}, as measured in terms of the unfaithfulness and Fisher matrix errors. In subsection~\ref{subsection_mismatch} we compute the unfaithfulness due to truncating the fiducial template in various ways (described below), to get an idea of the relative importance of various terms. Subsection~\ref{systematic vs statistical} identifies criteria under which systematic errors due to neglecting certain terms in the phase are smaller than statistical errors. 
Finally, subsection~\ref{statistical error subsection} discusses the convergence properties of the statistical errors.

\subsection{Unfaithfulness of truncated templates}                \label{subsection_mismatch}

We compute unfaithfulness distributions for the 1000 binaries in our catalog. We compare the eight truncated templates A-H defined in Sec.~\ref{subsec:truncate} (cf. Table~\ref{define templates}) against the fiducial template for three selected initial eccentricities $(e_0=0.04,\,0.07,\,0.1)$. A large value of the unfaithfulness indicates that the dropped term(s) are significant.

Our results are shown in Fig.~\ref{mismatch_histograms}. The histograms in the top panel show the effect of dropping terms of various orders in the initial eccentricity $e_0$ at 2PN order (i.e., we move ``horizontally'' along the top row Table~\ref{define templates}). As expected, the unfaithfulness decreases as we move from template A to template D: template D is the closest to the fiducial template.
Similarly, the middle panel shows the effect of dropping terms of various PN orders at order $e_0^6$  in the initial eccentricity (i.e., we move vertically along the right column of Table~\ref{define templates}). Again, as we move from template G to template D (thereby dropping all terms higher than the Newtonian, 1PN, 1.5PN and 2PN order terms, respectively) the unfaithfulness decreases.
Finally, the bottom panel corresponds to moving diagonally inwards along Table~\ref{define templates}, starting from the top-right corner. 
In each panel, the unfaithfulness gets larger as we increase $e_0$: this is expected, since all of these waveforms are small-eccentricity expansions.

An interesting exception is template A. In this case we are dropping a {\em circular} term of order 2PN and $e_0^0$, so the unfaithfulness is largely independent of $e_0$ (as it should be).
The small faithfulness ($M \sim 0.65$) of template A means that the 2PN-$e_0^0$ term is very important, and it is suggestive of the necessity to include higher PN orders for circular ($e_0^0$ order) templates. This is well known in the GW data analysis community, and it is indeed implemented in the $3.5$PN accurate circular template \texttt{TaylorF2}~\cite{creighton2012gravitational}.

\begin{table*}[htp]
\begin{center}
\centering
\begin{tabular}{| *5{>{\centering\arraybackslash}m{1.4in}|} @{}m{0pt}@{}}
\hline
 \centering   { 2PN  $x^2$ }     &     $3.7,~3.7,~3.7,~3.7$   &         $ \left( 100  ,  6.0   , 1.7  , 0.76 \right) \times 10^{2}$  &   $\left( 2300   , 8.3   ,    0.80  ,  0.17 \right) \times 10^{4}$  &   $  \left(2.1, 0.26,0.009,0.001   \right) \times 10^{8}$          \\ 
\hline
 \centering     {   1.5PN    ($x^{3/2}$)  }     &  $2.6,~2.6,~2.6,~2.6$    &    $170,~16,~7.0,~5.0$      &   $  \left( 6000, 23, 2.2, 0.58   \right)   \times 10^{2}$    &   $   \left( \infty, 130, 4.3,0.50   \right) \times 10^{4}    $     \\ 
\hline
\centering      {      1PN  ($x^1$)  }   &  $2.6,~2.6,~2.6,~2.6$       &  $28,~4.0,~3.1,~2.9$    &   $   \left( 500,1.8,0.21,0.081  \right) \times  10^2$   &     $   \left( 210000, 60, 2.0 ,0.22   \right) \times 10^{3}    $  \\ 
\hline
 \centering    {    Newtonian  ($x^0$) }  &    $2.7,~2.7,~2.7,~2.6$     &   $2.7,~2.6,~2.6,~2.6$       &  $   \left( 270, 3.7, 2.7,2.6  \right)$     &    $   \left( 17000, 4.0, 0.18 ,  0.050   \right) \times 10^{2}    $   \\ 
\hline
        & \centering        {     $e_0^0$      }       &        {      $e_0^2$    }         &       {       $e_0^4$    }       &         {      $e_0^6$    }      \\     
\hline
\end{tabular}
\end{center}
\caption{Maximum detection SNR -- computed using Eq.~(\ref{mismatch-SNR-2}) and the median faithfulness -- such that systematic errors are smaller than statistical errors when we drop the term corresponding to each box. Numbers correspond to $e_0= 0.01, 0.04, 0.07$ and $0.1$. $\infty$ means that the median unfaithfulness is zero within machine roundoff errors.
}
\label{SNR table}
\end{table*}

\begin{table*}[htp]
\begin{center}
\centering
\begin{tabular}{| *5{>{\centering\arraybackslash}m{1.4in}|} @{}m{0pt}@{}} 
\hline
 \centering   { 2PN    ($x^2$)  }      &     $-$   &         $ 0.16$  &   $0.21$  &   $0.30$          \\ 
\hline
 \centering     {   1.5PN  ($x^{3/2}$)  }     &  $-$    &    $0.03$      &   $0.12$    &   $0.24$      \\ 
\hline
\centering      {      1PN   ($x^1$) }   &  $-$      &  $0.008$    &   $0.06$   &      $0.18$ \\ 
\hline
 \centering    {    Newtonian  ($x^0$) }  &   $-$     &  $0.0007$    &  $0.015$     &    $0.05$     \\ 
\hline
        & \centering        {     $e_0^0$      }        &        {      $e_0^2$    }         &       {       $e_0^4$    }       &         {      $e_0^6$    }      \\     
\hline
\end{tabular}
\end{center}
\caption{Maximum $e_0$ such that detections with $\rho<25$ have systematic errors smaller than statistical errors, when terms corresponding to the respective cells in the table are dropped. A dash means that the mismatch is so low that systematic errors are larger than statistical errors for all $e_0$ in the range we consider.
}
\label{e0 table}
\end{table*}

\begin{figure*}[htp]
\begin{center}
\includegraphics[width=\textwidth, angle=0]{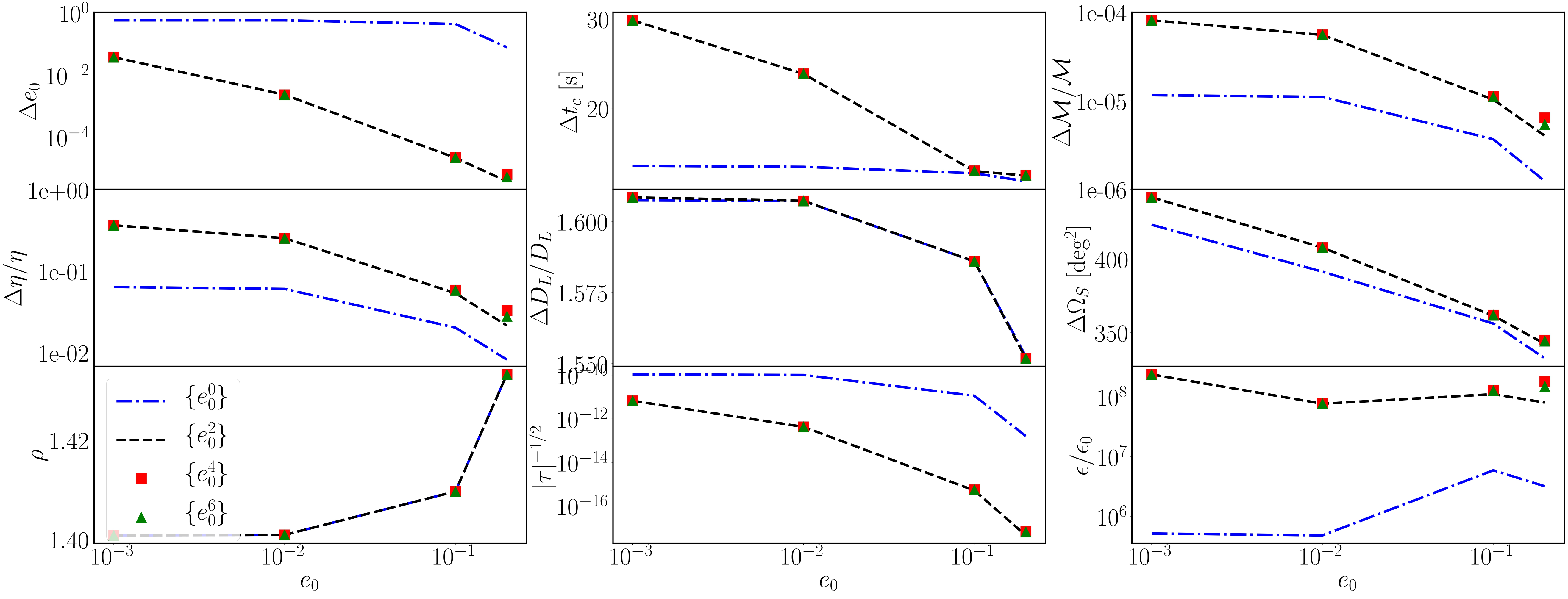}
\end{center}
\caption{
\label{fisher_errors}
Top two rows: statistical errors for the $ \left\lbrace e_0^0 \right\rbrace$, $ \left\lbrace e_0^2 \right\rbrace$, $ \left\lbrace e_0^4 \right\rbrace$ and $ \left\lbrace e_0^6 \right\rbrace$ (fiducial) template. The errors change going from the $ \left\lbrace e_0^0 \right\rbrace$ to the $\left\lbrace e_0^2 \right\rbrace$ template, but they are pretty much constant at higher orders in $e_0$.  Bottom: SNR (left); volume of the error ellipsoid $|\tau|^{-1/2}$ as given by the square root of the determinant of the inverse Fisher matrix (center); and correlation between the parameters, as measured by $\epsilon/\epsilon_0$ [see the discussion below Eq.~(\ref{mismatch-SNR-2})].}

\end{figure*}

\subsection{Systematic errors vs. statistical errors}    \label{systematic vs statistical}

In Fig.~\ref{mismatch_SNR} we study the conditions under which systematic errors are smaller than statistical errors for selected truncated templates. 
Each panel shows the unfaithfulness (left y-axis) and the SNR obtained when we replace the inequality in~(\ref{mismatch-SNR-2}) by an equality (right y-axis) of selected truncated templates as a function of $e_0$. Solid lines correspond to the median unfaithfulness (or SNR) over our sample of 1000 binaries. Shaded areas correspond to the 25th and 75th percentiles, so they give an idea of the spread in the data.

Consider, for example, a detection with SNR $\rho = 20$ and initial eccentricity $e_0 = 0.02$. If the corresponding point in one of these plots lies {\em below} the unfaithfulness-SNR curve of the corresponding template, then systematic errors are negligible with respect to statistical errors. The faithfulness by itself is not sufficient to decide whether systematic errors are negligible: we also need Eq.~(\ref{mismatch-SNR-2}) to determine the maximum SNR beyond which systematic errors dominate. Note also that it would be incorrect to use these plots for low SNRs, since large SNRs were assumed to derive Eq.~\eqref{mismatch-SNR-2}. 

Another way to read these plots is as follows. Suppose that we want to compute the posterior distribution for a detection with maximum likelihood corresponding to $e_0 = 2 \times 10^{-2}$, $\rho \sim 20$.
If we want systematic errors to be negligible with respect to statistical errors when we construct the posterior distribution, we can choose a template whose unfaithfulness-SNR curve at $e_0 \sim 2\times 10^{-2}$ (as shown in Fig.~\ref{mismatch_SNR}) gives $\rho>20$. For example, templates B, C and D satisfy this criterion, whereas template A does not. Let us remark once again that Fig.~\ref{mismatch_SNR} should not be trusted for low SNRs, therefore (for example) the unfaithfulness-SNR curve for template A cannot be trusted in this example.

Curves corresponding to low unfaithfulness (or high $\rho$) mean that the GW template will have negligible systematic errors (recall that a point must lie below the unfaithfulness-SNR curve for systematic errors to be negligible). Figure~\ref{mismatch_SNR} implies that systematic errors become negligible as we move from templates A to D, G to D and H to D, i.e. as we move towards templates which are closer to the fiducial template, and our mismodeling errors become smaller. Furthermore, the ratio of systematic to statistical errors becomes smaller as $e_0$ decreases ($\rho$ gets higher as $e_0$ decreases). An exception is template A, for which the dominant dropped term is independent of $e_0$.

When are systematic errors negligible with respect to statistical errors? Focus, for example, on template H in the bottom-left panel of Fig.~\ref{mismatch_SNR} and on the two templates in the bottom-right panel. For template H, systematic errors are negligible in the whole range $0<e_0 < 0.1$ when $\rho< 10^3$. For the $\left\lbrace  e_0^4 \right\rbrace$ template, the bottom-right panels shows that systematic errors become negligible in the range $e_0 < 6 \times 10^{-2}$ for SNRs below the blue curve. The corresponding range is smaller for the $\left\lbrace  e_0^2 \right\rbrace$ template. In these regions the truncated template can be used for parameter estimation to save computational time and the dominant errors are statistical. The convergence of statistical errors will be the topic of the next subsection.

Table~\ref{SNR table} -- built through the inequality~(\ref{mismatch-SNR-2}) -- shows the maximum detection SNR for which systematic errors are smaller than statistical errors for templates obtained by dropping the term corresponding to a given cell, and for selected values of $e_0=0.01,\,0.04,\,0.07,\,0.1$.
Similarly, Table~\ref{e0 table} shows the maximum $e_0$ such that detections with $\rho<25$ have systematic errors smaller than statistical errors when terms corresponding to the respective cells in the table are dropped.

\subsection{Convergence of statistical errors}      \label{statistical error subsection}


Unfaithfulness-SNR plots can be used to identify templates for which systematic errors are smaller than statistical errors. Given such a template, statistical errors can be computed (in the high-SNR limit) as the square root of the diagonal elements of the inverse Fisher matrix. We computed statistical errors for the 1000 binaries in our catalog using the $\left\lbrace  e_0^0\right\rbrace$, $\left\lbrace  e_0^2\right\rbrace$, $\left\lbrace  e_0^4\right\rbrace $ and $\left\lbrace  e_0^6\right\rbrace$ templates (where the last one is the fiducial template).\footnote{Similar calculations were performed in~\cite{Nishizawa2016} for two templates: a circular template and a template at leading order in $e_0$ with different amplitudes.} In Fig.~\ref{fisher_errors} we plot median statistical errors for these four templates.

Statistical errors change going from template $\left\lbrace  e_0^0\right\rbrace$ to template $\left\lbrace  e_0^2\right\rbrace$, but then they plateau. This convergence of statistical errors was not observed in~\cite{Nishizawa2016}, where errors were computed only for the circular and ${\cal{O}}(e_0^2)$-accurate templates.

The error on the eccentricity $\Delta e_0$ decreases when we go from a circular template $\{e_0^0\}$ to the $\{e_0^2\}$ template, but it is roughly constant as we increase the order of the $e_0$ expansion. This is because the phase of the circular template $\{e_0^0\}$ is independent of $e_0$, so all information comes from the amplitude alone, leading to large errors. The errors $\Delta t_c$, $\Delta\Omega_S$, and $\Delta \ln D_L$ decrease mildly (within a factor of two) as $e_0$ increases for all four templates: these are all extrinsic parameters for which measurement information comes largely from the motion of the detector, which is not significantly affected by the template we use.
The mass errors $\Delta \mathcal{M}/\mathcal{M}$ and $\Delta \eta/\eta$ are underestimated by a factor of 5-10 when we use the circular template $\{e_0^0\}$, and they are largely the same for all eccentric templates $\{e_0^n\}$ with $n=2,\,4,\,6$; in other words, the simplest eccentric template $\{e_0^2\}$ already contains enough information to estimate mass measurement errors. Note also that most errors (with the exception of $\Delta e_0$) vary by at most factor of 2 as functions of $e_0$.

\begin{figure*}[htp]
\begin{center}
\includegraphics[width=\textwidth, angle=0]{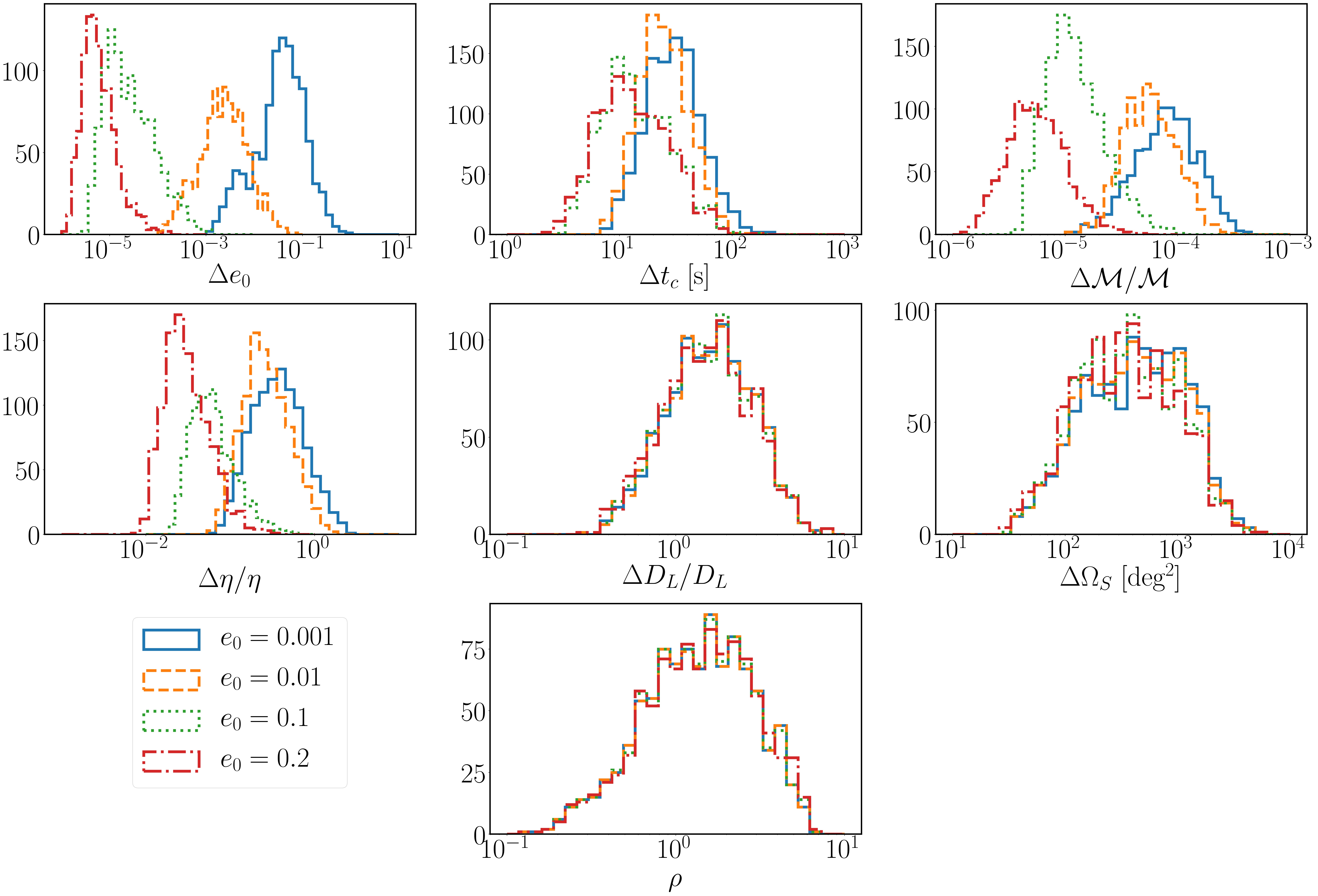}
\end{center}
\caption{
\label{fisher_errors2}
Statistical error histograms for the fiducial template and four selected values of $e_0$.
} 
\end{figure*}

The bottom row of Fig.~\ref{fisher_errors} addresses the question: how do statistical errors change as we increase the order of the $e_0$ expansion in the phase?
The bottom central panel shows that $|\tau|^{-1/2}$ (the volume of the error ellipsoid, as given by the square root of the determinant of the inverse Fisher matrix) decreases going from the $ \left\lbrace e_0^0  \right\rbrace$ to the $\left\lbrace e_0^2  \right\rbrace$ template: the 11-dimensional error ellipsoid shrinks (i.e., statistical errors decrease) with more accurate templates.
This is in apparent contradiction with previous plots, showing that many errors on individual parameters increase. The solution to this apparent paradox (as shown in the bottom-right panel) has to do with correlations between parameters, as measured by $\epsilon/\epsilon_0$ -- see the discussion below Eq.~(\ref{mismatch-SNR-2}): this quantity\footnote{We slightly modify the definition of the correlation factor to stand for the product of $\Delta {\cal{M}}/{\cal{M}}$, $\Delta {\cal{\eta}}/{\cal{\eta}}$, $\Delta t_c, \Delta \phi_c$, $\Delta {\cal{D}}_L/{\cal{D}}_L$, $\Delta \Omega_S$, $\Delta \Omega_L$, $\Delta \beta$ and $\vert \tau \vert^{1/2}$.
Note that $\vert \tau \vert$ is equal to the reciprocal of the parameter $\epsilon_0$ introduced in Sec.~\ref{dabg}, thus quantifying an overall measure of the statistical errors.} increases by a factor of $10^2 - 10^3$ going from the $\left\lbrace  e_0^0\right\rbrace$ template to the $\left\lbrace  e_0^2\right\rbrace$ template, and then remains roughly constant.
Recall that all plots refer to a fixed redshift $z=0.1$ ($D_L = 447.8$~Mpc), but errors scale linearly with $D_L$ in the large-SNR limit.

In Fig.~\ref{fisher_errors2} we plot histograms of the statistical errors for all 1000 binaries using our fiducial template. These histograms essentially confirm the conclusions of~\cite{Nishizawa2016}. We can measure the initial eccentricity as long as $\Delta e_0<e_0$: this is true for most binaries when $e_0>0.1$. If $e_0 = 0.01$, $\Delta e_0 < e_0$ for about $90 \%$ of the binaries in our sample.

Before closing this section, we would like to mention that setting $\beta=0$ (as was done in \cite{Nishizawa2016}, thereby reducing the number of parameters from 11 to 10) results in a decrease of $\Delta \Omega_S$ by a factor of $\sim 10$. This suggests that the parameter $\beta$ is correlated with sky location parameters, and that setting it to zero can lead to an underestimation of those errors.

\section{Conclusions}
\label{conclusions}

We studied the convergence of the frequency domain 2PN-$e_0^6$ order accurate ``fiducial'' GW templates for compact eccentric binaries of~\cite{Tanay2016}. We built truncated templates by dropping certain terms, and assessed the importance of those terms by computing the unfaithfulness (Fig.~\ref{mismatch_histograms}). Dropping most terms leads to unfaithfulness $<0.02$ for $e_0 < 0.1$. The terms that produce the largest unfaithfulness when dropped are the 2PN$-e_0^0$ and 0PN$-e_0^6$ terms; extensions at 0PN$-e_0^n$ with $n>6$~\cite{YABW09} and $m$PN$-e_0^0$ with $m>2$~\cite{LR_LB} (e.g., the \texttt{TaylorF2} approximant) are already available in the literature.

We then investigated the conditions under which truncated templates produce systematic errors which are smaller than statistical errors (Fig.~\ref{mismatch_SNR}). This helps us to identify ``fast templates'' that can be used for parameter estimation considering only statistical errors. 

In Fig.~\ref{fisher_errors} we studied the convergence of statistical errors. Statistical errors converge very quickly, and they do not change much as long as we include terms of order $e_0^2$ in the phasing. More accurate templates yield larger statistical errors than the $ \left\lbrace  e_0^0\right\rbrace$ template for most parameters, with the exception of the error $\Delta e_0$ on the initial eccentricity. However the error ellipsoid shrinks as we increase the order of the $e_0$ expansion: indeed, statistical errors for most of the individual parameters increase because of the larger correlations between parameters. Figure~\ref{fisher_errors2} shows statistical errors for the fiducial template, and it confirms the main conclusions of Ref.~\cite{Nishizawa2016} (which used slightly different templates).

Several extensions of this work are possible and necessary. An important limitation of our study is that we kept only the three leading-order harmonics in our templates; future work should further explore the convergence of $\tilde{h}(f)$ [Eq.~(\ref{GW strain})] as the number of harmonics changes.  Our analysis is specific to stellar-origin BH binaries observed with LISA, but similar work should be done for second- and third-generation Earth-based detectors and using other templates (see e.g.~\cite{DGI, konigs}). There are ongoing efforts to extend our ``fiducial templates''~\cite{Tanay2016} to 1PN order in amplitude and 3PN order in phase, including the effects of periastron advance \cite{STiwari2019}. As soon as these templates are available, an extension of our analysis can be used to assess the relative significance of PN amplitude corrections with respect to phase corrections. It will also be important and useful to extend our study to Fourier-domain templates accurate at 3PN and valid for large eccentricities, which are currently under development~\cite{2018CQGravityYunes,Moore-19}. 

\section*{Acknowledgments}
We would like to thank Leo Stein and Kaze W.K. Wong for discussions and suggestions.
E.B. is supported by NSF Grant No. PHY-1841464, NSF Grant No. AST-1841358, NSF-XSEDE Grant No. PHY-090003, and NASA ATP Grant No. 17-ATP17-0225. 
This work has received funding from the European Union’s Horizon 2020 research and innovation programme under the Marie Skłodowska-Curie grant agreement No. 690904. 
Computational work was performed at the Mississippi Center for Supercomputing Research (MCSR) and at the Maryland Advanced Research Computing Center (MARCC).
A.N. is supported by JSPS KAKENHI Grants No. JP17H06358 and No. JP18H04581.

\appendix

\section{Oscillatory cross-terms in the Fisher matrix}
\label{appendix_A}

As discussed in Sec.~\ref{sectionI}, the cross terms in the integrand of Eq.~(\ref{cfm1}) are highly oscillatory, and thus can be neglected.
This can be understood analytically as follows. Let us first truncate the templates of~\cite{Tanay2016} at leading order in both the PN parameter $x$ and the initial eccentricity $e_0$ (i.e., we consider circular templates). We first decompose the template into its first six harmonics:
\begin{align}
  \tilde{h}(f) &= \tilde{h}_1(f) + \tilde{h}_2(f) + \tilde{h}_3(f)+  \tilde{h}_4(f) + \tilde{h}_5(f) + \tilde{h}_6(f), 
\end{align}
and then we decompose each harmonic into an amplitude and a phase to get
\begin{widetext}
\begin{align}
 \tilde{h}_j(f) &= \sum_{j} A_j(f) e^{i \Psi_j(f)}           ,      \\    
 \Psi_j(f) &=  2 \pi f t_c - j \phi_c + \frac{3}{128 \eta}  {\left(\frac{G m_z \pi f}{c^3}  \right)}^{-5/3} \left(\frac{j}{2}\right)^{8/3} - \frac{\pi}{4}  .
\end{align}

By Eq.~(\ref{cfm1}), the Fisher matrix elements involve derivatives of the template with respect to the parameters:
\begin{align}
 \frac{\partial \tilde{h}_j(f)}{\partial p_A} &= \left[ \frac{\partial A_j}{\partial p_A} + i A_j \frac{\partial \Psi_j}{\partial p_A} \right] e^{i \Psi_j}.
\end{align}
and they can be broken down into a sum of integrals of the form
\begin{align}
 \Re \int_{f_{\rm min}}^{f_{\rm max}} \frac{1}{S_n(f)}\frac{\partial \tilde{h}_{j_1}(f)}{\partial p_A} \frac{\partial \tilde{h}_{j_2}(f)^*}{\partial p_B} df &=  \Re \int_{f_{\rm min}}^{f_{\rm max}} \frac{1}{S_n(f)} \left[ \frac{\partial A_{j_1}}{\partial p_A} + i A_{j_1} \frac{\partial \Psi_{j_1}}{\partial p_A} \right] \left[ \frac{\partial A_{j_2}}{\partial p_B} - i A_{j_2} \frac{\partial \Psi_{j_2}}{\partial p_B} \right] e^{i \left( \Psi_{j_1} - \Psi_{j_2} \right)} df, 
\end{align}
as a result of the coupling of different harmonics $(j_1,\,j_2)$ and different parameters $(p_A,\,p_B)$. The phase term in the exponential reads
\begin{align}
 \Psi_{j_1} - \Psi_{j_2} &= (j_2 - j_1) \phi_c + \frac{3}{128\eta}{\left(\frac{G m_z \pi f}{c^3}  \right)}^{-5/3} \left[ \left(\frac{j_1}{2} \right)^{8/3} - \left(\frac{j_2}{2} \right)^{8/3} \right].
\end{align}
\end{widetext}
Therefore the integrand has a rapidly oscillatory phase $\Delta \Psi \propto f^{-5/3}$ whenever $j_1 \neq j_2$.
For diagonal elements of the Fisher matrix ($p_A = p_B$) terms with $j_1 = j_2$ always exist, and those integrals dominate. For certain off-diagonal terms (e.g. the $\ln D_L$-$\phi_c$ term), the integrals with $j_1 = j_2$ exactly vanish. However, the natural scale for these terms is set by the corresponding diagonal elements of the Fisher matrix, and the integrals with $j_1 \neq j_2$ can still be neglected.

\section{Beam pattern functions and other quantities appearing in the templates}      \label{beam_pattern_appendix}

In this appendix, for completeness, we define certain quantities appearing in the GW strain via [Eqs.~(\ref{GW strain}), (\ref{GW strain2}), (\ref{GW strain3}) and (\ref{GW strain4})]. It is well known that certain combinations of trigonometric functions of the eccentric anomaly $u$ of a binary with eccentricity $e$ can be written as Fourier-Bessel series~\cite{YABW09,maggiore2008gravitational}:
\begin{align}
\frac{\sin u}{1-e \cos u}  =  2   \sum_{k=1}^{\infty}  J_k'(k e) \sin k l,    \\
\frac{\cos u}{1-e \cos u}  =  \frac{2}{e}   \sum_{k=1}^{\infty}  J_k(k e) \cos k l,
\end{align}
where $J_k$ denotes Bessel functions of the first kind
\begin{align}
J_k(x)  =   \sum_{n=0}^{\infty}      \frac{(-1)^m}{n! ~ \Gamma(n+k+1)  }   \left(\frac{x}{2}  \right)  ^{2 n+k},
\end{align}
and $\Gamma$ is the Gamma function. When combined with the well-known relations involving the orbital phase $\phi$ of a Keplerian orbit
\begin{align}
\cos \phi  &=   \frac{\cos u  - e}{1 - e \cos u}      , \\
\sin \phi  &=    (1-e^2)^{1/2} \frac{\sin u }{1 - e \cos u}   , 
\end{align} 
the equations above yield
\begin{align}
\cos \phi   &=     -e + \frac{2}{e} (1- e^2)      \sum_{k=1}^{\infty}  J_k (k e) \cos k l       \label{phi eqn1}         \\                 
\sin \phi   &=  (1-e^2)^{1/2}       \sum_{k=1}^{\infty}  (J_{k-1} (k e) - J_{k+1} (k e)) \sin k l   .             \label{phi eqn2}
\end{align}

Following~\cite{YABW09}, the plus and cross polarizations can be written as 
\begin{align}
& h_{+}  =  - \frac{G^2 \mu}{c^4 p D_L}\left[  \left(  2 \cos (2 \phi - 2  \beta ) +\frac{5 e}{2} \cos(\phi  -  2 \beta)               \right.\right.\nonumber\\ 
  &\qquad \left.\left.{}   + \frac{e}{2} \cos(3 \phi - 2 \beta)      +  e^2 \cos (2 \beta) \right)  (1 +  \cos^2 {\iota})        \right.\nonumber\\  
  &\qquad \left.{}       +    (  e \cos \phi + e^2) \sin^2 {\iota}        \right]    ,     \\
& h_{\times}   =       - \frac{G^2 \mu}{c^4 p D_L}   \left[   4 \sin(2 \phi - 2  \beta) + 5 e \sin(\phi - 2  \beta )         \right.\nonumber\\  
  &\qquad \left.{}      + e \sin (3 \phi -  2 \beta ) - 2 e^2 \sin (2 \beta)    \right] \cos {\iota},    
\end{align}
where $\mu = m_1 m_2/(m_1+m_2)$ and $p$ (the semilatus rectum of the orbit) is related to the orbital angular frequency $\omega$ via
\begin{align}
\omega  = (m_1+m_2)^{1/2}    \left(  \frac{p}{1-e^2} \right)^{-3/2}.
\end{align}
where the inclination angle ${\iota}$ is defined by $\cos \iota = \hat{L}  \cdot \hat{N}$~\cite{MP99,Yagi2010}, where $\hat{L}$ and $\hat{N}$ are unit vectors in the direction of the orbital angular momentum and in the direction of the source, respectively.
By plugging Eqs.~(\ref{phi eqn1}) and (\ref{phi eqn2}) into the expressions for $h_{+}$ and $h_{\times}$ above, we get
\begin{align}
h_{+, \times}   =    -  \frac{G^2  {\cal{M}}  }{c^4 D_L} ( {\cal{M}}  \omega  )^{2/3}   \sum_{j = 1}^{\infty}  \left[    C_{+,\times}^{j} \cos(j l)  +  S_{+,\times}^{j} \sin(j l)   \right].
\end{align}
The quantities $C_{+,\times}^{j} $ and $S_{+,\times}^{j}$ read~\cite{YABW09}
\begin{align}
C_{+}^{1}  &=    e \left(-\frac{3 c_{2
   \beta }}{2}-\frac{3}{2} c_{2 \beta }
   c_i^2+s_i^2\right)  \nonumber   \\   &    + \frac{1}{24} e^3 \left(16 c_{2 \beta
   }+16 c_{2 \beta } c_i^2-3
   s_i^2\right)  ,      \\   
S_{+}^{1}  &=     -\frac{3}{2} e
   \left(c_i^2+1\right) s_{2 \beta }   +    \frac{23}{24} e^3 \left(c_i^2+1\right)
   s_{2 \beta },       \\
C_{\times}^{1}  &= 3 ~e ~c_i s_{2 \beta }-\frac{4}{3} e^3 c_i
   s_{2 \beta } ,      \\
S_{\times}^{1}  &=  -3 ~e~
   c_{2 \beta } c_i     +  \frac{23}{12} e^3 c_{2 \beta } c_i    ,     \\
C_{+}^{2}  &=     2 \left(c_{2 \beta }+c_{2 \beta } c_i^2\right)    +e^2 \left(-5 c_{2
   \beta }-5 c_{2 \beta } c_i^2+s_i^2\right)   \nonumber \\ 
   &  +  \frac{1}{12} e^4 \left(33 c_{2 \beta }+33 c_{2 \beta } c_i^2-4 s_i^2\right) ,       \\   
S_{+}^{2}  &= 2
   \left(c_i^2+1\right) s_{2 \beta } -5 e^2 \left(c_i^2+1\right) s_{2 \beta }  \nonumber \\
   &  +    3 e^4 \left(c_i^2+1\right) s_{2 \beta }       ,       \\
C_{\times}^{2}  &=      -4 c_i s_{2 \beta }   +10 e^2 c_i s_{2 \beta }    -\frac{11}{2} e^4 c_i s_{2 \beta }   ,      \\
S_{\times}^{2}  &=  4 c_{2 \beta } c_i   -10 e^2 c_{2 \beta } c_i +   6 e^4 c_{2 \beta } c_i ,     \\
C_{+}^{3}  &= \frac{9}{2} e \left(c_{2 \beta }+c_{2 \beta } c_i^2\right)   \nonumber \\
&-\frac{9}{16} e^3 \left(19 c_{2 \beta
   }+19 c_{2 \beta } c_i^2-2 s_i^2\right),        \\   
S_{+}^{3}  &=      \frac{9}{2} e \left(c_i^2+1\right) s_{2 \beta }-\frac{171}{16} e^3 \left(c_i^2+1\right) s_{2
   \beta }   ,       \\
C_{\times}^{3}  &=  -9 e c_i s_{2 \beta }  +  \frac{171}{8} e^3 c_i s_{2 \beta }    ,      \\
S_{\times}^{3}  &=  9 e c_{2 \beta } c_i-\frac{171}{8} e^3 c_{2 \beta } c_i ,     \\
C_{+}^{4}  &=   8 e^2 \left(c_{2 \beta }+c_i^2 c_{2 \beta }\right) \nonumber \\
&  +e^4 \left(-20 c_{2 \beta }-20 c_i^2 c_{2
   \beta }+\frac{4 s_i^2}{3}\right) ,      \\   
S_{+}^{4}  &=   8 e^2 \left(s_{2 \beta }+c_i^2 s_{2 \beta }\right)  -20 e^4 \left(s_{2 \beta }+c_i^2 s_{2 \beta
   }\right),       \\
C_{\times}^{4}  &=     -16 e^2 c_i s_{2 \beta }+40 e^4 c_i s_{2 \beta } ,      \\
S_{\times}^{4}  &=  16 e^2 c_i c_{2 \beta }-40 e^4 c_i c_{2 \beta }   ,     \\
C_{+}^{5}  &=     \frac{625}{48} e^3 \left(c_{2 \beta }+c_{2 \beta } c_i^2\right) ,       \\   
S_{+}^{5}  &= \frac{625}{48} e^3 \left(c_i^2 s_{2 \beta }+s_{2 \beta }\right)     ,       \\
C_{\times}^{5}  &=     -\frac{625}{24} e^3 c_i s_{2 \beta }  ,      \\
S_{\times}^{5}  &=  \frac{625}{24} e^3 c_{2 \beta } c_i ,     \\
C_{+}^{6}  &= \frac{81}{4} e^4 \left(c_{2 \beta }+c_{2 \beta } c_i^2\right),        \\   
S_{+}^{6}  &=   \frac{81}{4} e^4 \left(c_i^2 s_{2 \beta }+s_{2 \beta }\right)    ,       \\
C_{\times}^{6}  &=    -\frac{81}{2} e^4 c_i s_{2 \beta }    ,      \\
S_{\times}^{6}  &=  \frac{81}{2} e^4 c_{2 \beta } c_i ,    
\end{align}
where $c_{2 \beta} = \cos {2 \beta},~s_{2 \beta} = \sin {2 \beta},~c_i= \cos {\iota}$ and $s_i= \sin {\iota}$. The GW strain at the detector is the linear combination $h(t) = F_{+} h_+  +  F_{\times}  h_{\times}$,
and its Fourier transform is given by Eq.~(\ref{GW strain}). The beam pattern functions $F_{+}$ and $F_{\times}$ can be found, e.g., in~\cite{Berti:2004bd,Yagi2010}.

\bibliography{Amybib}
\end{document}